# Vision-based interface for grasping intention detection and grip selection : towards intuitive upper-limb assistive devices

E. Moullet[a,b]*, F. Bailly[a], J. Carpentier[b], and C. Azevedo-Coste[a]

[a]CAMIN, INRIA centre d'Université Côte d'Azur, Université de Montpellier, Montpellier, France ;
[b]WILLOW, INRIA Paris, Département Informatique de l'ENS, Paris, France

## 1. Introduction

Numerous pathologies may affect upper-limb movements (quadriplegia, stroke, amputation…). In particular, grasping is crucial for many daily activities, and its impairment considerably impacts quality of life and autonomy. Attempts to restore this function may rely on various approaches and devices (functional electrical stimulation, exoskeletons, prosthesis…), and all face two great challenges. First, targeted objects' characteristics are highly variable in shape, weight, and texture, demanding high flexibility from the assistive device in controlling its degrees of freedom (DOFs). Second, individuals with disabilities have a significantly reduced number of DOFs available to express their intent, and current human-machine interfaces provide limited inputs for controlling devices (Azevedo et al. 2022; Jiang and Farina 2014). Current approaches (i.e. myoelectric control) rely on state machines to alternatively set the user's DOF as the control input for a given (potentially synergetic) movement elicited by the device. Yet these command modalities often exert considerable cognitive loads on users and lack controllability and intuitiveness in daily life.

In this work, we propose a novel user interface for grasping movement control in which the user delegates the grasping task decisions to the device, only moving their (potentially prosthetic) hand toward the targeted object. Time before impact is estimated for the device to open or prepare the hand in time for the upcoming grip, which is automatically selected among a set of grips achievable by the assistive device.

## 2. Methods

### 2.1 Apparatus and computer vision tools

Required information for assisting an ongoing grasping task is the following: 1) hand position and orientation; 2) object position and orientation; 3) object nature (including shape and potentially weight and texture). Regarding hand pose estimation, markerless motion capture applying convolutional neural network (CNN) to RGB frames is a most active field of research that has consistently been narrowing the accuracy gap that used to separate them from optoelectronic systems, all while relying on less expensive and cumbersome hardware. In the meantime, object pose estimation most active research also relies on CNNs applied to RGB frames. Thus, an OAK-D S2 (Luxonis) stereoscopic RGB camera was chosen as a data acquisition sensor, prioritizing cost, ease of use, installation and universality, as it may be easily mounted on a wheelchair or a hat. Hand pose estimation was performed using Google's MediaPipe (Zhang et al. 2020), leveraging stereoscopic vision for depth estimation. Object identification and pose estimation were achieved using CosyPose (Labbé et al. 2020), a multi-object 6D pose estimator trained on a set of objects with known 3D models.

### 2.2 Grasping intention detection

Hand and objects' pose were estimated and represented in a 3D virtual scene for grasping intention detection. Multiple rays formed a virtual cone in the direction of the hand's velocity, with a length and diameter embedding scalar velocity. Ray tracing was used to determine if the user targeted an object by comparing the number of rays intercepting with the object's 3D mesh to a threshold.

### 2.2 Time before impact estimation

We defined the expected area of impact as the mesh triangle intercepted by a ray defined as the barycenter of the intercepting rays. Time before impact was evaluated by integrating the extrapolated velocity of the hand up to the area of impact.

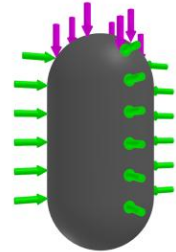

Figure 1. Illustration of the adaptive grip selection: different approach movements (represented as arrows embedding $\Delta_{\square\square\square}$ and $\square_{\square\square\square\square}$) yield either a palmar grip (green) or a pinch grip (purple)

### 2.3 Grip selection

We chose to select a grip among a predefined set rather than generate a grip on the flight for two reasons: 1) some assistive devices cannot (yet) achieve independent, continuous control over the hand (Azevedo et al. 2022); 2) predefined grips are more likely to be correctly anticipated and exploited by the user. We took inspiration from the concept of documented objects (Dalibard et al. 2010) to select the optimal grip for a given object and a foreseen vector of impact. A 4-dimensional table (illustrated in Fig.1) was constituted for each object of the used set in which a grip is attributed to every tuple formed by the mesh triangle representing the expected area of impact $\Delta_{imp}$ and the hand's normed velocity $v_{hand}$.

## 3. Results and discussion

In a preliminary study, a participant was instructed to grasp randomly selected objects with a randomly selected grip. Video recordings of the task were analyzed offline to fine-tune the parameters of our grasping intention detection algorithm and build the grip selection table.

Our ongoing study aims at quantifying the real-time performances of our algorithm regarding accuracy and time of impact estimation, which are crucial for practical use. Fig. 2 shows a snapshot of the video feed of a grasping task on the upper part and the corresponding virtual scene on the lower part. Object 000028 was correctly detected as the target, while object 000023 was ignored. Area of impact was estimated from the ray intercepts figured as red dots, and a palmar grip was accordingly selected.

An upcoming study in which participants will wear an assistive (FES or prosthetic) device will assess the usability of the proposed approach.

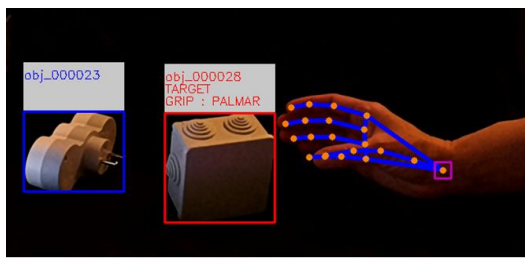

Figure 2. Visualization of the adaptive grasping assistance algorithm

Although this proof of concept used CosyPose, which is restricted to the variety of objects it was trained on, recent work allows for pose estimation of objects unseen during training but with known 3D models. Nonetheless, the fact that our grip selection table was built specifically for our set of objects may present challenges in real-time scenarios where object discovery is required. Despite these limitations, the proposed approach shows potential in controlled environments with known objects, such as home settings, where it can significantly reduce the cognitive load for individuals with upper-limb pathologies.

Additionally, it should be noted that the proposed approach is agnostic of the hand and object pose estimation tool. Any precise and fast enough pose estimation solution could be utilized, allowing for flexibility in choosing appropriate technologies based on specific requirements and constraints.

Finally, while only the 3D position of the palm was used, exploiting the whole hand pose estimation and prolonging the tracking after the object is grasped would allow, for instance, increases in the grip strength if the object is slipping as a substitution to missing sensory feedback.

## 4. Conclusions

Grasping intention detection and automatic adaptive grip selection were achieved, requiring minimally complex sensors. This proof of concept paves the way for assistive device control modes lightening the cognitive load of impaired people in known environments.

**\*Corresponding author. Email: etienne.moullet@inria.fr**